\documentclass[a4paper,nodate]{sab} 

\usepackage{graphicx} 
\usepackage{txfonts,natbib} 
\bibpunct{(}{)}{;}{a}{}{,} 
\usepackage{hyperref}
\usepackage{url}
\usepackage[T1]{fontenc} 
\usepackage{t1enc} 
\renewcommand\ackname{Acknowledgements.} 
\renewcommand\andname{\&}
\renewcommand\keywordname{Keywords.} 
\renewcommand\figureptname{Figure}
\renewcommand\tableptname{Table}

\renewcommand\noteaddname{Note added to proofs.}
\renewcommand\andname{\&}

\idline{Boletim da Sociedade Astron\^omica Brasileira, {\bf 31}, no. 1, XX-XX}{3}
\begin{document} 
\title{Information technology \& astronomical data in Brazil} 
\subtitle{Perspectives and proposals} 

\author{U. Barres de Almeida\inst{1}, A. Krone-Martins\inst{2}, M. Diaz\inst{3}, J. D. do Nascimento\inst{4}, W. Leo\inst{5}, R. R. Rosa\inst{6} \and R. K. Saito\inst{7}} 
\institute{Centro Brasileiro de Pesquisas F\'{i}sicas (CBPF), Rio de Janeiro, Brasil \email{ulisses@cbpf.br} \and University of California, Irvine-CA, USA \email{algolkm@gmail.com} \and Universidade de S\~{a}o Paulo (IAG-USP), S\~{a}o Paulo, Brasil \email{marcos.diaz@iag.usp.br} \and Universidade Federal do Rio Grande do Norte (UFRN), Natal, Brasil \email{jddonascimento@gmail.com} \and Laborat\'{o}rio Nacional de Computa\c{c}\~{a}o Cient\'{i}fica (LNCC), Petr\'{o}polis - RJ, Brasil \email{wagner@lncc.br} \and Instituto Nacional de Pesquisas Espaciais (INPE), S\~{a}o Jos\'{e} dos Campos-SP, Brasil \email{rrrosa.inpe@gmail.com}  \and Universidade Federal de Santa Catarina (UFSC), Florian\'{o}polis-SC, Brasil \email{saito@astro.ufsc.br}}
\date{Received } 

\Abstract {The Commission on Science and Information Technology (CTCI) of the Brazilian Astronomical Society (SAB) is tasked with assisting the Society on issues of astronomical data management, from its handling and the management of data centres and networks, to technical aspects of the archiving, storage and dissemination of data. In this paper we present a summary of the results of a survey recently conducted by the Commission to diagnose the status of several data-related issues within the Brazilian astronomical community, as well as some proposals derived therefrom.}{A Comiss\~{a}o em Tecnologia e Ci\^{e}ncia da Informa\c{c}\~{a}o (CTCI) da Sociedade Astron\^{o}mica Brasileira (SAB) tem como objetivo principal assistir a Sociedade em temas ligados \`{a} gest\~{a}o de dados astron\^{o}micos, desde o seu utilizo e a gest\~{a}o de centros de dados e redes, at\'{e} os aspectos t\'{e}cnicos relacionados ao arquivo, processamento e dissemina\c{c}\~{a}o dos dados. Neste artigo, apresentamos um resumo dos resultados de uma consulta recentemente conduzida pela Comiss\~{a}o para diagnosticar o atual estado de diversos temas relacionados \`{a} gest\~{a}o de dados na comunidade astron\^{o}mica Nacional, bem como algumas propostas da\'{i} resultantes.}



\maketitle 

\section{Rationale}
%

The primary function of the Committee on Science and Information Technology (CTCI) of the Brazilian Astronomical Society (SAB) is to advise and assist SAB, and other interested parties, on all issues involving Information Technology and Data Science, in order to provide a pragmatic and efficient development of Brazilian astronomy. The committee discusses, with open participation to all SAB members, and proposes what their practical actions should be (based on the demands to be identified) in order to achieve the desired goals. 

As a starting point of its activities, the Commission has targeted the following set of actions: 

\begin{itemize}
\item Outline the national direction and set national priorities in astronomical research in relation to the needs of RD\&I in information technology and data science, where topics such as Big Data and e-Science stand out;
\item Encourage and promote inter-institutional initiatives that undertake computer-assisted RD\&I activities in Astronomy, in a way that reinforces cooperation and collaboration between national and international science and technology institutions and personnel;
\item Promote and maintain cooperation with other Information Technology and Data Science-related committees and associations;
\item During its initial years of work, the main objective of the Commission is to present a technical white paper on the directions and strategies in Data Science and Information Technology (DS\&IT) for Astronomy, to be considered by SAB in favour of the community it represents. The present document is a first working draft, resulting from an extensive survey conducted within SAB on the topic. 
\end{itemize}

\section{Background}
Astronomy and astrophysics are data-driven sciences. Whenever a new observation window is opened or a new instrument is available, the base of astronomical knowledge is constituted, in the first instance, by means of surveys that seek to make a complete and impartial census of the sky. Today, such surveys are conducted in all bands of the electromagnetic spectrum, from radio waves to gamma rays, and more recently also in the field of the so-called multi-messengers, which include cosmic rays, neutrinos and gravitational waves.

Since the pioneer \emph{Carte du Ciel} - probably the first major international scientific collaboration on a global scale -- that began in the late 19th century, and which generated a total of 22,000 photographic plates from all over the sky -- these efforts have become increasingly ambitious~\citep{jones}. To get an idea of the exponential progression on data accumulation, the total volume of data acquired by typical surveys from a decade ago is equivalent to just one night of observation by several among the main instruments available today~\citep{djorgovski}.

In less than a decade this dynamic will undergo yet another a qualitative leap. If since the turn of the Century the progression in data volume can be roughly described by a ''Moore's law'',  the production of data in Astronomy in the next decade will grow by an estimated two to three orders of magnitude. As an example, we can cite, at high energies, the Cherenkov Telescope Array (CTA) observatory, which will produce about 12 PB per year (or 30 TB a day --~\cite{lammana}), similarly to the future Rubin Observatory Legacy of Space and Time Survey (LSST), in optical, which will produce circa 15 PB per year of data (see \url{https://www.lsst.org/about/dm}); Both are still an order of magnitude below the amount of data expected to be generated by the Square Kilometer Array (SKA), which will generate close to 300 PB a year of data at radio waves (see~\url{https://www.skatelescope.org/sdp/} -- equivalent to almost 30 TB per year per professional astronomer in the world! Such data will make Astronomy one of the largest data producers among all scientific disciplines -- comparable to CERN's Large Hadron Collider (LHC).

Due to the transient nature of many of the astrophysical sources, astronomers must also be concerned with the temporal dimension of the observations in order to characterise the variability of the objects. The scales of variability observed are very diverse, and can evolve over long periods, of years or decades, or be as brief as a fraction of a second, in the case of extreme, usually serendipitous, events. This gives us a multi-dimensional view of the range of modern astronomical observations, where the space-time dimension must add to the multi-band and multi-messenger observations to offer a complete view of the physical processes operating in the cosmos (e.g.,~\cite{charles},~\cite{obrien}).

As such, the observational, processing, archival and data management efforts demanded by modern astronomy are immense, and are likely to become increasingly complex. From a technical point of view, we can mention: (i) the need for a coordination of observations between different instruments in the study of variable sources, which presupposes a global and efficient alert and information exchange / integration system for the monitoring of variable events~\citep{williams}; (ii) a large capacity for real-time processing or daily \emph{in situ} data reduction, as well as the capacity of transmission of large amounts of information from remote locations to data centres worldwide(~\cite{zhang},~\cite{price}); (iii) and finally the keeping of large files and associated technology for data access, on which heavy demands on usability and interoperability weight on, to allow for efficacious multi-temporal and multi-band studies~\citep{genova}.

From a methodological point of view, we are further faced, within this context of ''big data'' science, with the challenge of managing large international collaborations. Here, work is highly specialised, and there is a growing need for the use of technologies and methods based on artificial intelligence to inspect data that becomes progressively too bulky for direct or supervised analysis by specialists to take place at first instance~\citep{kremer}. Likewise, within this dynamics, the requirements on computer simulations capabilities, in order to accompany the evolution in details achieved by observations, is also very significant~\citep{taffoni}.

\section{The survey}

\begin{table}[t]
\centering
\begin{tabular}{|c|c|}
\hline
Data Volume & Respondents \\
\hline
< 10 TB & 60\% \\
10-100 TB & 31\% \\
100 TB - 1 PB & 9\% \\
> 1 PB & 0\% \\
\hline
\end{tabular}
\caption{Volume of annual data generation by observational groups and facilities.}
\label{tab:datavolume}
\end{table}

In the next section we will present an overview of the status of information technology and scientific astronomical data in Brazil, on the basis of the results of a survey on the subject conducted by the CTCI Commission in early 2019. Following that, a set of short-term future perspectives and proposals for the National astronomical community are also discussed. The survey reached the entire community of members of SAB and was responded by circa 120 members, which, in our view, suffices to guarantee representativeness.

The survey was prepared through a series of discussions and activities by the Commission, including a Workshop on Data Science and Information Technology organised at the National Institute of Space Research (INPE) in 2017\footnote{\url{http://www.lac.inpe.br/wtcia2017/}}, as well as presentations and discussions during the Annual Meetings of SAB.

The survey was divided in two major areas, respectively focused on Information Technology and Computation (Part I) and Data Science (Part II). Within Part I, the following topics were present: algorithm development and programming language; data processing and the use of computational resources; big data handling / storage and data volume. 

In part II the survey concentrated in topics associated to: type and volume of data produced by the Brazilian community;  data publication and sharing policy; data usage and access; open data and multi-band /-messenger research. 

For this second part of the survey, respondents were asked to identify themselves according to the their main band / messenger of expertise, so that we could identify possible particularities within different areas of observational astronomy. The respondents were also asked to identify their research as majorly resulting on the provision or publication of new data, or chiefly based on the use of data provided by other groups. Those who were, at some level, responsible for facilities, groups or initiatives / collaborations associated to observations or data provision of interest to the broader community, were also asked to identify themselves as such. For the remainder, the survey was entirely anonymous.

\section{Results and diagnosis} 

\begin{table*}[t]
\centering
\begin{tabular}{|l|c|c|c|c|}
\hline
\textbf{Cataloguing} & Fully & Partially & None (Proprietary) & None (Lack of Resources) \\
\textbf{of primary data} & 44\% & 17\% & 25\% & 14\% \\
\hline
\textbf{Storage Mode} & Closed Repository & Local Open DB & Local VO-based & External Open DB\\
\textbf{of primary data} & 74\% & 11\% & 9\% &  6\% \\
\hline
\textbf{Reuse of the} & Over 50\% & Under 50\% & Not significant & Don't know \\
\textbf{primary data} & 17\% & 19\% & 11\% & 53\% \\
\hline
\end{tabular}
\caption{Data storage and archiving practices by the Brazilian astronomical community and statistics on the re-use of data.}
\label{tab:catalogs}
\end{table*}

\subsection{Part I: Information Technology and Computing} 
\subsubsection{Algorithms and programming languages}
In the first part of the survey, the first question was related to the development of algorithms and programming languages. It was identified that about 60\% of the community develops its own algorithms and computer programmes, and that the vast majority does so without the help or the collaboration of professional programmers or IT professionals. The most used computer languages for coding are Python, FORTRAN (still resisting the test of time, possibly due to the ongoing use of large astrophysical legacy codes) and C++. Python, as expected, is the language whose usage is growing the fastest. 

The low rate of collaboration with IT professionals in the development of code signals to a low level of production of software and data science products destined to external groups or communities;  it also attests to a weak level of interaction with researchers or teams from applied computation disciplines. In modern astronomy, where computation and data science are becoming ever more central to the daily practices of the academic research, both these practices should be reinforced. Likewise, the community could benefit from the formation of professionals with a focused expertise on programming and astronomical data science. The development of more collaborations and interfaces with researchers and professionals from applied computation would also be to the general benefit of the national astronomical community.

\subsubsection{Research bottlenecks}
We also surveyed which were the most important bottlenecks in terms of computational infrastructure for astronomical research. The issues manifested by the community spanned several infrastructure itens, but specially in relation to the availability and access to an adequate computer processing capacity, particularly for multi-core parallel computing or GPU-based computing. We also detected, nevertheless, that the community sub-utilises some of the most advanced resources available in the Country -- e.g., only 5\% of respondents have used, or plan to use in the short-term, the Santos Dumont supercomputer (the largest in Latin America), and just under 10\% recourses to cloud computing or remote open clusters available throughout the Country for their every-day work. This underutilisation of facilities are likely the result of a combination of three factors: lack of an adequate training and support to the use advanced facilities; lack of knowledge about the open computer infrastructures available in the Country; and finally, lack of a coherent and centralised policy and administration for research computer infrastructures which facilitates accessibility. 

A significant fraction of the sampled community (33\%) also manifested a lack of the necessary technical knowledge for big data manipulation and for advanced data science techniques. In this regard, when surveyed about the volume of (observational or computer-generated / simulation) data typically dealt with over the course of an year, 70\% of the community responded that it typically generates / processes with under 1 TB of data, and only a very small minority operates with more than 100 TB over the course of one year of research. Such data volumes can be considered small by current international standards, and set to grow fast in the coming decade, which implies that the infrastructure and training issues identified above may become majors hinders for astronomical research in the medium term if not actively mitigated.

In answering to these points, the Commission encourages the advancing of policies that supports the use of powerful, distributed computing facilities. A survey of the open computational resources available from the different institutions around the Country, and the creation of a single reference point or directory from which these resources could be accessed is highly desirable to promote its use; such task could be conducted by an appropriate National institution such as LNA or SAB itself. Finally, the use of funding for building shared, especially cloud computing infrastructures, accessible to the entire national astronomical community, stands out to this Commission as a prime avenue to help solving several of the problems raised, especially by our colleagues outside the big centres, the so-called Sao Paulo - Rio axis.

\begin{table*}[t]
\centering
\resizebox{\textwidth}{!}{
\begin{tabular}{|l|c|c|c|c|c|}
\hline
\textbf{Origin of main data} & P.I.-owned & Collaboration & Open raw data & Open final products & Publications \\
\textbf{used in research} & 38\% & 23\% & 21\% & 13\% & 5\% \\
\hline
\textbf{Data destination} & Closed & Collaboration & Raw data available & Science products available & \\
\textbf{after scientific publication} & 31\% & 40\% & 19\% &  10\% & \\
\hline
\textbf{Origin of multi-band} & P.I.-owned & Collaboration-owned & Open raw data & Open high-level products & Publications \\
\textbf{used in research } & 34\% & 26\% & 21\% & 16\% & 4\% \\
\hline
\end{tabular}}
\caption{Data usage in research and data publication after first scientific utilization by groups in the Brazilian astronomical community. Specific information is also given about the origin of multi-band data used in research.}
\label{tab:usage}
\end{table*}

\subsection{Part II: Data Science}
Data science will become an increasingly important part of the daily research activities of the professional astronomer, and the results of the survey conducted by this Commission show that it is necessary to include basic topics of data science in the formal training of young astronomers, in order to keep pace with future global developments. This could be most efficiently achieved by partnering with computation and informatics departments or institutes for formal degrees and curricula, or through the organisation of specific and regular training schools. In fact, the very low usage by the community of Application Programming Interfaces (APIs), which is at the basis of client-server communication and data applications (by circa 15\% of the respondents), signals to restricted data science and big data activity in the national astronomical community.

\subsubsection{Data provision and archives}
For the continuation of this second part of the survey, we split the community into three groups, according to their research profile: "data users", "data providers" or "observatory / instrument manager or responsible". For those groups associated to data production or management, we have surveyed what are the typical data volumes dealt with over an year period. The results are shown in table~\ref{tab:datavolume} and show numbers that are still generally small -- probably owing to the National deficit in modern astronomical observatories and a general lack of organised data centres -- or the difficulties with computer infrastructure resources.

Table~\ref{tab:catalogs} shows how the generated data is archived or catalogued by the groups or researchers that are the major data producers within the community. Although a fair amount (over 60\%) of all data generated by the different groups is catalogued online, still the vast majority (71\%) is done so in local, closed repositories; only a very small amount of catalogues (20\%) is open to external access, and an almost negligible 9\% is stored in Virtual Observatory (VO)-compatible formats. The obvious result of this scenario is that there is very low record of data re-usability by the community, which is probably very small, with only an estimated 10-20\% of all data produced being reused by other researchers than the original PIs.  Such low statistics on the open archiving and reuse of the data generated by the community represent a critical issue for research impact, reproducibility of results, and accountability of the use of public resources.

\subsubsection{Data utilisation and multi-band astronomy}
Table~\ref{tab:usage} looks at the mirror situation to the previous analysis, inquiring about the data access and usage practices by groups or researchers that are mostly ''data users'' (as opposed to generators of data). When inquired about the origin of the data used in their research, most users (38\%) answered to be the PI of the observations or project. A significant fraction use data generated internally by the collaborations to which they are members (23\%), and only 1/3 critically uses data that is open and produced by other groups for their research. This picture shows that data generated by observational groups in Brazil are often not made available online for future reuse by others, which reflects in the practice of using mostly PI data for one's own research (a trend that is globally reversing fast). A demonstration of this behaviour is that, when it comes to multi-band research, 60\% uses data for which they or their collaborations are the PIs. Such low proportions go against the trend of increased multi-band / multi-messenger research. 

The use and sharing of high-level (final) data products is a yet less diffuse practice within the community, with negative implications for data transparency: only 10\% publish their high-level data products in open databases for general use by others. Although this is still the rule in most astronomical communities around the world, the next generation of large observatories and facilities, and most of all, the large surveys, is set to change this.

\begin{figure}[h]
\caption{Statistics of the use of VO services and tools and of accessibility to open data bases by the Brazilian astronomical community.}
\centering
\includegraphics[width=0.5\textwidth]{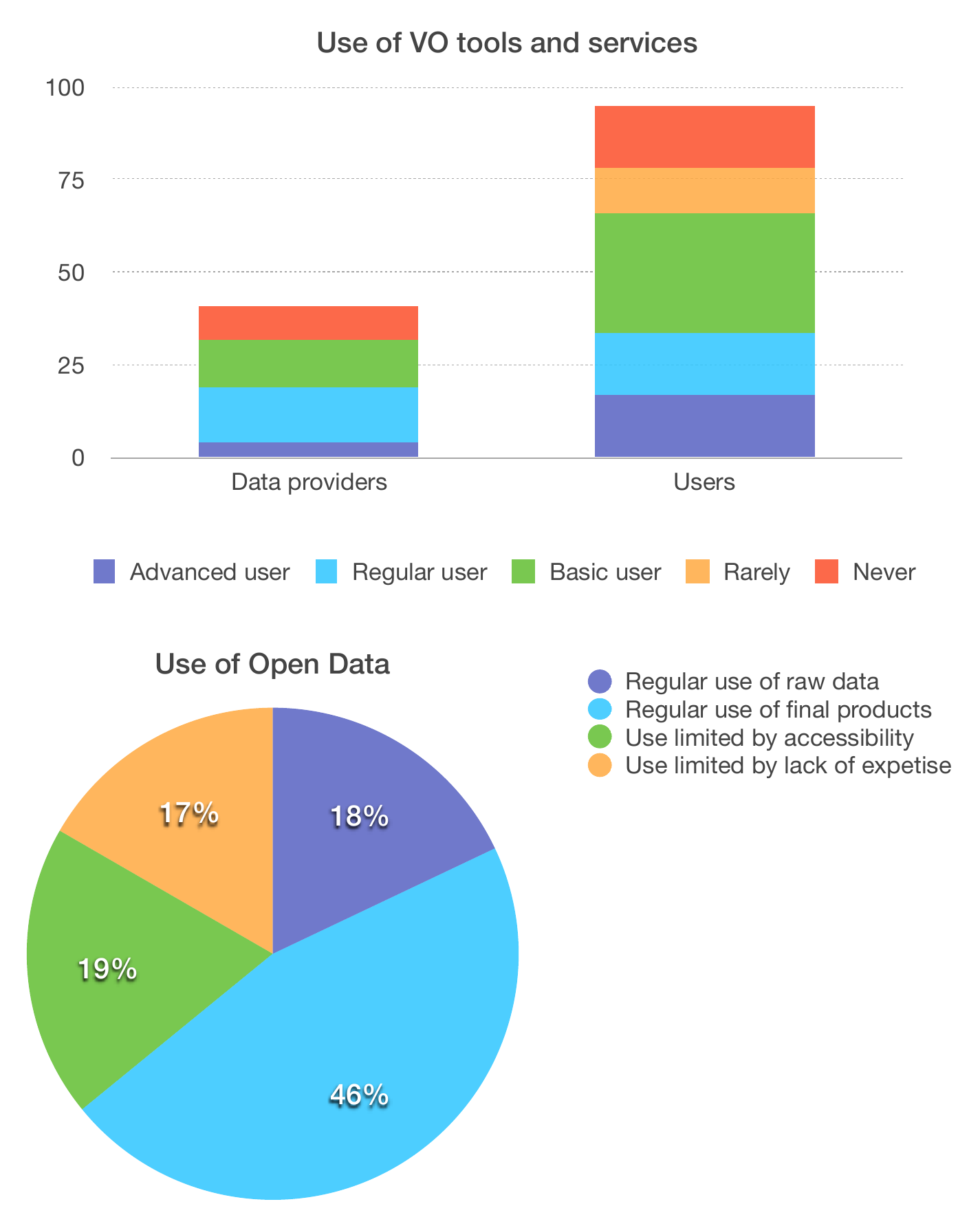}
\label{fig:OpenVO}
\end{figure}

Finally, the survey asked about the usage of open data and VO tools by the community, as shown in figure~\ref{fig:OpenVO}. The results show that circa 30\% among data providers are advanced or regular users of VO tools and services for publication of their data, a statistics similar to that of final users. Nevertheless, the vast majority of the community do not recourse to VO tools in their research, or does so only at a basic level. When viewed against the next plot, which shows that almost 70\% of respondents are regular users of open data (raw or final products alike), this shows that the astronomy research community could benefit from more formal training on the use of VO tools and services (or similar data science resources).

\section{Summary and proposals} 

As a first outcome of this survey, firstly presented in these proceedings, we wish to stimulate some discussion and actions to try to address the aspects we have identified as most critical for the short / medium-term future of the national astronomical community.

\begin{itemize}
\item Firstly, the results of the survey reinforce the need for \textbf{training young astronomers on topics of computer programming and in particular data science}, and to stimulate and broaden the \textbf{interfaces and interactions of the astronomical community with the applied computer science communities};
\item A national / institutional debate is necessary to understand the best way to \textbf{improve accessibility to computational resources throughout the Country}, and especially outside the ''large centres''. As a first approach, we suggest that a possible, most effective solution, is to stimulate the efficient sharing of large computational infrastructures already available in various institutions, and the \textbf{integration of resources nationally through cloud computing solutions} -- similarly, for example, to what is being done in Europe with the European Science Cloud (ESC);
\item Finally, \textbf{Virtual Observatory (VO) tools and its associated databases and services} are an integral part of the astronomical infrastructure, with ever growing importance in the medium-term future. Specially in the current context of limited resources for new astronomical research infrastructure and observatories in Brazil, the effective use of such tools, which allows for an impactful publication of data products, and wide access to quality open data, is critical for the near-term future of astronomical research in the Country, and general further internationalisation of the Brazilian astronomical research.
\end{itemize}

\section{Future steps} 
The Commission on Science and Information Technology (CTCI) has been charged by the Brazilian Astronomical Society (SAB) with the task of undertaking a study of the situation of ''Data Science and Information Technology'' in Brazilian Astronomy. The results of the survey presented here serve as a first diagnosis to guide discussions and future actions.

Following from that, and with the help of experts and institutional actors within the community, we wish to start an in-depth study of the three main topics listed in the previous section -- IT and data science education in astronomy, national resources for astronomical computation, virtual observatory and data resources -- in order to look for solutions and directives that may guide future actions, as well as national policies and funding decisions.

Such is a complex community work, for which the engagement of SAB and of the professional astronomers and students, as well as of research institutions, universities, and government / funding agencies is crucial.  This year we aim to take this discussion forward, with the objective of producing white papers focusing on guiding the policy on the three main topics listed.

\begin{acknowledgements} The authors are grateful for the technical support of Paulo H. Barchi in collecting information published in this article. This paper is the result of the work of the Commission on Science and Information Technology (CTCI) of the Brazilian Astronomical Society (SAB) and presents the results of a survey conducted within the entire community of Brazilian astronomers, whose response and support we gratefully acknowledge here. \end{acknowledgements} 

\end{document}